\documentstyle[psfig]{l-aa}
\newcommand{\be}{\begin{equation}}
\newcommand{\en}{\end{equation}}
\begin{document}
\def\ltsima{$\; \buildrel < \over \sim \;$}
\def\lsim{\lower.5ex\hbox{\ltsima}}
\def\gtsima{$\; \buildrel > \over \sim \;$}
\def\gsim{\lower.5ex\hbox{\gtsima}}
\def\spose#1{\hbox to 0pt{#1\hss}}
\def\approxlt{\mathrel{\spose{\lower 3pt\hbox{$\sim$}}
        \raise 2.0pt\hbox{$<$}}}
\def\approxgt{\mathrel{\spose{\lower 3pt\hbox{$\sim$}}
        \raise 2.0pt\hbox{$>$}}}
\def\deg {^\circ}
\def\mdot {\dot M}
\def\kms {$\sim$km$\sim$s$^{-1}$}
\def\gs {$\sim$g$\sim$s$^{-1}$}
\def\ergs {$\sim$erg$\sim$s$^{-1}$}
\def\cmtre {$\sim$cm$^{-3}$}\def\nupa{\vfill\eject\noindent}
\def\der#1#2{{d #1 \over d #2}}
\def\l#1{\lambda_{#1}}
\def\grb{$\gamma$-ray burst}
\def\grbs{$\gamma$-ray bursts}
\def\rosat{{\sl ROSAT} }
\def\cmdue {cm$^{-2}$}
\def\gcm {$\sim$g$\sim$cm$^{-3}$}
\def\rsole{$\sim$R_{\odot}}
\def\msole{$\sim$M_{\odot}}
\def\aa #1 #2 {A\&A, {#1}, #2}
\def\mon #1 #2 {MNRAS, {#1}, #2}
\def\apj #1 #2 {ApJ, {#1}, #2}
\def\nat #1 #2 {Nature, {#1}, #2}
\def\pasj #1 #2 {PASJ, {#1}, #2}
\newfont{\mc}{cmcsc10 scaled\magstep2}
\newfont{\cmc}{cmcsc10 scaled\magstep1}
\newcommand{\bc}{\begin{center}}
\newcommand{\ec}{\end{center}}

\title{The X--ray Diffuse Emission from the Galactic Center.}
\author{L.~Sidoli\inst{1,\,2} \& S.~Mereghetti\inst{1}}

\institute{
{Istituto di Fisica Cosmica ``G. Occhialini" -- C.N.R., Via Bassini 15, I-20133 Milano,
Italy;  \\ e-mail: (sidoli, sandro)@ifctr.mi.cnr.it}
\and
{Dipartimento di Fisica, Universit\`a di Milano, Sez. Astrofisica, Via Celoria 16, I-20133
Milano, Italy}
}

\thesaurus{09.19.2, 10.03.1, 13.25.3, 13.25.4}

\offprints{L.~Sidoli}

\date{Received 13 July 1999 / Accepted 2 August 1999}

\maketitle
\label{sampout}

\begin{abstract}

A long BeppoSAX observation of the Galactic Center region 
shows that the spectrum of the diffuse X--ray emission 
from the SgrA Complex can be described with the sum of two
thermal plasma models with temperatures of $\sim$0.6 keV 
and $\sim$8-9 keV.
The spatial distribution of the diffuse emission is energy dependent.
While the hard X--ray emission has a nearly symmetric distribution 
elongated along the Galactic Plane, the soft X--rays  
(E $<$ 5 keV) are remarkably well correlated with the triangular
radio halo of SgrA East. The parameters derived for the soft 
component support the interpretation of the SgrA East shell
in terms of an evolved supernova remnant.


\end{abstract}
 
\section{Introduction}

The   diffuse X--ray emission from the Galactic Center (GC) region was 
imaged for the first time with  
the Einstein Observatory   in the 0.5--4 keV band (Watson et al. 1981) 
and later studied in more detail with several  satellites
(Kawai et al. 1988, Skinner et al. 1987, 
Koyama et al. 1989, 1996).
The real nature of this emission is still an open issue. 
Though at least part
of it can be due to the integrated emission from weak unresolved 
sources (Watson et al. 1981; Zane, Turolla \& Treves 1996), there is 
evidence that a hot plasma, 
responsible for the observed emission lines,
permeates the GC region. This was indicated by 
the Ginga discovery of a strong iron line at 6.7 keV   
with an  equivalent width of
$EW=600\pm$70 eV (Koyama et al. 1989). More recently, the ASCA satellite confirmed 
the presence of a   hot plasma (kT$\sim10$~keV)  and 
discovered   another  diffuse component  in the 6.4 keV Fe-line 
with an asymmetric spatial distribution with respect to 
the GC and well  correlated with the distribution
of the giant molecular clouds (Koyama et al. 1996).
The data on the diffuse X--ray emission 
reported here were obtained during a survey
of the GC performed with the BeppoSAX Narrow Field Instruments
(Sidoli et al. 1998). The results on the point sources
have been   reported elsewhere (Sidoli et al. 1999a, 1999b).

\section{The Galactic Center environment}

\begin{figure}[!ht]
\vskip 0.truecm
\centerline{\psfig{figure=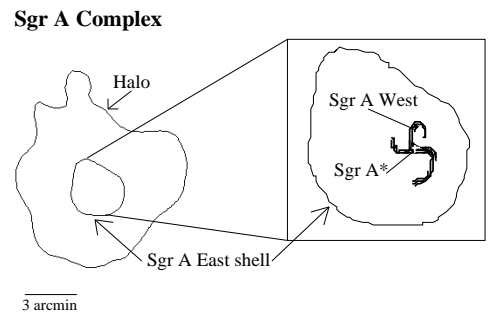,width=6.5cm} {\hfil}}
\caption{A schematic view of the main radio structures 
in the SgrA Complex. North is to the top, East to the left. 1 arcmin corresponds to about 2.5 pc
at 8.5 kpc.}
\label{sgra}
\end{figure}

Before presenting our results, we briefly describe
the most important radio emitting sources present in the region
within $\sim8'$ from SgrA* covered by our 
observation (see Morris \& Serabyn 1996 for 
a recent review). These extended sources   are collectively known as the  
Sgr~A~Complex (Fig.\ref{sgra}).

The most peculiar object is the compact radio 
source SgrA* (Balick \& Brown 1974). 
It has a radio luminosity of $\sim2\times10^{34}$~ergs~s$^{-1}$ and 
a flux density rising as S$_{\nu}\propto\nu^{1/3}$ between a 
low frequency turnover at $\nu\sim$800~MHz
and a cut--off at 2000--4000 GHz 
(Mezger, Duschl \& Zylka 1996). Lo et al. (1998) 
measured an intrinsic size of less than 5.4$\times10^{13}$~cm 
with multi--wavelength VLBA observations.  From the proper motion
of the nearby IR stars a dark mass of 2.6$\times10^{6}$~M$_{\odot}$ has
been deduced to reside   within a 
central volume of 10$^{-6}$~pc$^{3}$ (Ghez et al. 1998). Sgr~A* is 
generally considered to be a super--massive black hole due to its   unique 
radio spectrum  (see e.g. Krichbaum et al. 1998), 
its compactness, 
its location at the GC 
(Yusef-Zadeh, Choate \& Cotton 1999) and its 
low proper motion ($\leq$40 km~s$^{-1}$, Reid et al. 1999). 
The  high energy emission of SgrA* 
is surprisingly low, corresponding 
to a 2--10 keV luminosity of 
only $\sim$10$^{35}$~ergs~s$^{-1}$ 
(Koyama et al. 1996, Sidoli et al. 1999b), well
below the Eddington limit for a 2 million solar masses black hole.

SgrA* is embedded in the ionized gas of 
SgrA~West, an HII region
with a characteristic ``mini-spiral" shape of thermal radio 
emission. With  an angular
extent of $\sim1'$,  SgrA~West is the innermost and  ionized part 
of the  
``Circumnuclear Disk",  a  structure of molecular 
gas extending from 1.7 pc 
up to about 7 pc from the GC and rotating 
around it (Marshall, Lasenby \& Harris 1995).

Proceeding outward, we find the $2'-3'$ shell 
of SgrA~East (Eckers et al. 1983),
which is characterized by a non--thermal radio emission 
with  spectral index  $\alpha=-0.64$. It is centered about 
$50''$ offset of SgrA*
and is probably  located 
behind SgrA~West (Davies et al 1976, Pedlar et al. 1989). Its 
synchrotron radio emission and its shell morphology  
were explained in terms
of a supernova remnant. Mezger et al. (1989) discovered  
a dust ring compressed by the SgrA~East shell, 
possibly indicating that this shell originated from  an extremely energetic 
explosion (4$\times10^{52}$~ergs), maybe associated 
with SgrA*, occurring
inside a dense molecular environment 
(10$^{4}$~cm$^{-3}$). Other alternative explanations
are the simultaneous explosion of about 40 
supernovae, or a single SN explosion
inside a medium with a much lower density. In this case 
the dust shell could be due to the stellar wind 
from the normal supernova 
progenitor (Mezger et al. 1989). Another possibility 
is the tidal disruption of a star by 
the central supermassive black
hole (Khokhlov \& Melia 1996).

A triangular shaped non--thermal halo ($\sim$20 pc in diameter) 
surrounds in projection the SgrA~East shell 
(Yusef-Zadeh \& Morris 1987), but it is  probably   not
physically related with it (see 
Pedlar et al. 1989 for a detailed discussion). 
Its origin is unknown. It could be centered 
at the SgrA* position  and  related with
the high energy activity at the GC 
(Yusef-Zadeh et al. 1997, Melia et al. 1998). The total 
energy in relativistic particles estimated from
radio observations  is about 
5$\times10^{50}$~ergs~s$^{-1}$.
This value, together with its non--thermal radio spectrum and its 
size, suggest the  possibility that it is an evolved supernova
remnant  (Pedlar et al. 1989).

\section{Observations and Data Analysis}

Our results were obtained with the  Medium  Energy Concentrator 
Spectrometer    
(MECS, Boella et al. 1997) that provides images in 
the  $\sim$ 1.3--10 keV energy range  
over a nearly circular field of view with  $\sim28'$ radius.
The MECS has a good angular resolution
(50\% power radius of  $\sim75''$ at 6 keV, on-axis) and 
a moderate energy resolution 
(FWHM $\sim$8.5$\sqrt{6/{\rm E_{keV}}}$\%).  In order to 
exploit   the best spatial and spectral 
resolution of the detector, and to avoid the 
artifacts due to the  absorbing 
strongbacks of the MECS entrance window, we only considered
the inner region of the field of view (radius $\sim8'$). 

For the spectral analysis we have used the MECS effective area values 
appropriate  for extended sources. These were properly derived by 
convolving a flat  
surface brightness distribution  with the   energy and position 
dependent vignetting and Point Spread Function (PSF).

All the spectra have been 
corrected for the  the instrumental  background by subtracting
a spectrum obtained from MECS observations of the dark Earth.
This procedure does not account for the subtraction of the
cosmic X--ray background (CXB). 
Indeed, the   subtraction of the MECS standard background obtained
from high galactic latitude pointings would overestimate the  
CXB contribution, which is highly absorbed 
in the GC region. 
The CXB contribution has been included  as a known additive  
component in the spectral fits with an absorbing 
column density of 8$\times10^{22}$~cm$^{-2}$ and spectral parameters 
fixed at the values determined
with the MECS (photon index = 1.44, 
F(1 keV)=12.9~phot~cm$^{-2}$~s$^{-1}$~sr$^{-1}$~keV$^{-1}$; S. Molendi,
private communication).

\section{X--rays from the Sgr~A~Complex}

\begin{figure}[!ht]
\vskip -0 truecm
\centerline{\psfig{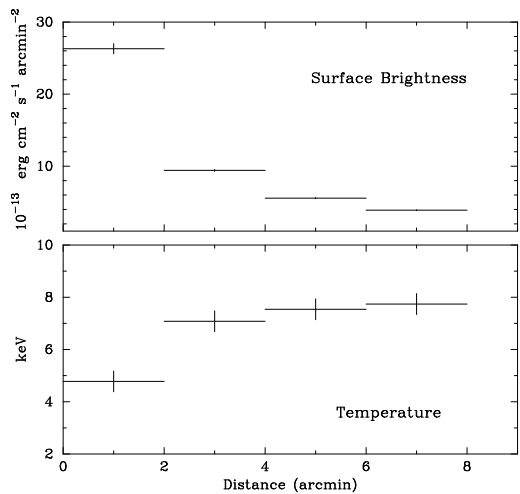} {\hfil}}
\caption{Surface brightness and temperature in concentric regions around the GC.}
\label{fsb}
\end{figure}

The location of SgrA* was imaged on August 24-26, 1997
for a net exposure time of 99.5 ks. 
In order to study the surface brightness and 
temperature profile of the diffuse emission, 
we extracted the  MECS   counts  from  four 
annular regions with different inner and outer radii
($0'-2'$,$2'-4'$,$4'-6'$,$6'-8'$)  centered on SgrA*.
All these spectra  contain several 
emission lines, with the K-lines from iron (E$\sim6.7$~keV) 
and sulfur (E$\sim2.4$~keV) particularly bright. We fitted
them with a single temperature plasma model
(MEKAL in XSPEC v.10.00), deriving the 
surface brigthness and temperature profiles 
shown in Fig.\ref{fsb}. A radial spectral 
variation is evident: while the 
$\sim7-8$ keV temperature  in the three external regions
is almost constant, the emission from the inner $2'$  is significantly
softer. This is probably due to the 
presence, in addition to the diffuse 
emission, of  further contributions    from the bright point sources
present near the GC (Maeda et al. 1996,  Predehl \& Tr\"umper 1994,
Sidoli et al. 1999b).

The fit with a single temperature thermal   model 
does not account for all the emission 
lines at low energies (in particular
the intense emission from sulfur at E$\sim$2.4 keV),
nor for an excess  around 6.4 keV, due to the
presence of emission of fluorescent origin. A  gaussian line 
added at 6.4 keV accounts for these residuals.
While this line is present in all
the spectra extracted from $2'$ to $8'$, it seems to
be absent within $2'$ (EW$\leq$12eV). This is consistent with 
the  6.4 keV map produced with ASCA (Maeda \& Koyama 1996)
where a strong peak is visible in the region of the
Radio Arc, NE to the GC. Indeed the EW of this line increases towards
the outer regions 
($EW_{2'-4'}=80-100$ eV,$EW_{4'-6'}=120-130$ eV, $EW_{6'-8'}=140-150$ eV).
The  MECS spectra 
extracted from two semi--annular regions at  NE and
SW of the GC, confirm this interpretation: the EWs of the 
6.4 keV line   are $\sim100\pm{30}$ eV (NE sector; errors
are at 90\% confidence level) and $\sim40\pm{20}$ eV (SW sector).

Since the temperature profile does not
show evidence of strong spectral variations in the region 
from $2'$ to $8'$,   we analysed the overall spectrum by 
extracting all the counts from this larger 
circular corona. We  started by fitting a thermal bremsstrahlung  
plus three gaussian  lines  at $\sim$1.8, 2.4  and 
6.7 keV (Si, S and Fe respectively). Their estimated 
equivalent widths are about
120, 190 and 1000 eV. 
The best fit value for the bremsstrahlung 
temperature is $\sim$13 keV. This temperature 
is too high to be consistent with the presence of the 
low energy emission lines (Kaneda et al. 1997, Fig. 2b). 
A possible explanation is a multi-temperature plasma. Thus we 
fitted the spectrum with two thermal emission plasma models
(two ``MEKAL" in XSPEC).  Our best fit parameters ($\chi^2$=1.29, 368 d.o.f)
are N$_{H}$=$7.93^{+0.09}_{-0.36}\times10^{22}$~cm$^{-2}$, T$_{1}=0.57^{+0.02}_{-0.05}$~keV   
and T$_{2}=8.69^{+0.26}_{-0.41}$~keV. A gaussian added at 6.4 keV 
gives an EW$\sim$120~eV. The total unabsorbed flux (2--10 keV) is 
$1.71^{+0.05}_{-0.08}\times10^{-10}$ ergs~\cmdue~s$^{-1}$,  about one third of which is  
contributed by the soft component.

\begin{figure*}[!ht]
\vskip 0.2truecm
\centerline{\psfig{figure=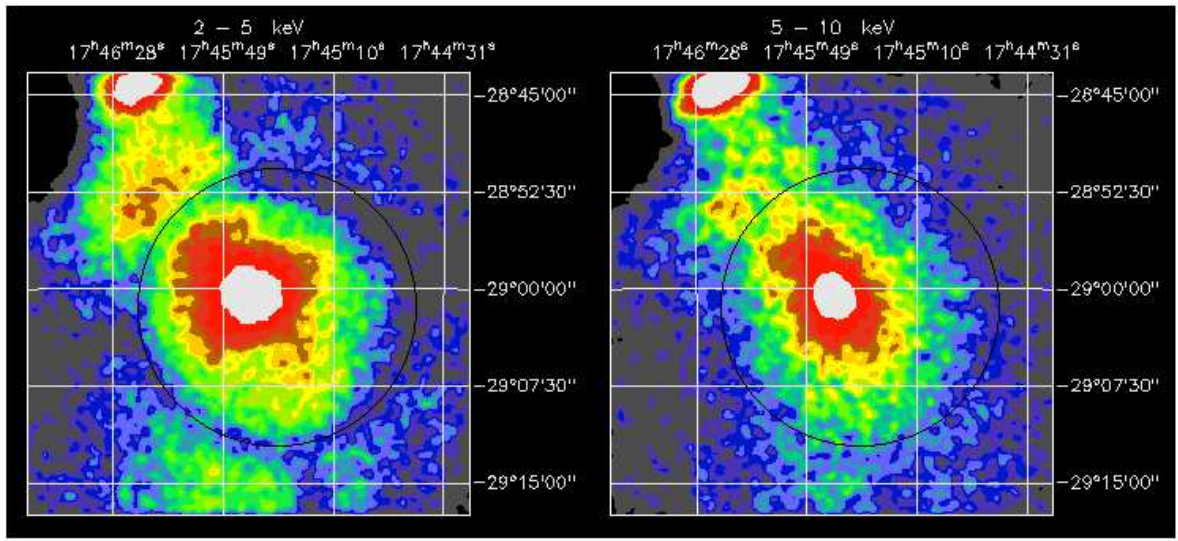,width=16.0cm} {\hfil}}
\caption{2--5 keV (left) and 5--10 keV (right)  emission from the SgrA~Complex. Both 
images have been smoothed with
a gaussian with FWHM=1 arcmin.   The strong source in the NE corner is 
1E1743.1--2843 (Cremonesi et al. 1999). The low surface brightness in correspondence 
of the circle is an 
instrumental effect due to the absorption in the detector window support structure.}
\label{softhard}
\end{figure*}

To  study the spatial distribution of the diffuse 
emission, we  extracted images in different
energy bands (Fig.\ref{softhard}).
The   images corresponding   to the
2--5 keV and   7--10 keV  ranges  show significantly different
distributions of the diffuse emission.

Both the soft and hard emissions are peaked at the GC position, but they
have   different spatial extents. 
In particular, the softer emission has a nearly triangular
shape, with a rather sharp decrease in the south-western side,
which is absent in the hard X--ray map.
To study the spatial distribution of the iron line we have also 
extracted an  image in the 5.5-7.5 keV   band and 
subtracted from this map the continuum emission interpolated
from the contiguous energies.  
The resulting iron line image shows a   spatial
distribution elongated along the galactic
plane, very similar to that of the  hard X--ray   map.

\section{Discussion}

\begin{figure}[!ht]
\vskip  -2.5truecm
\centerline{\psfig{figure=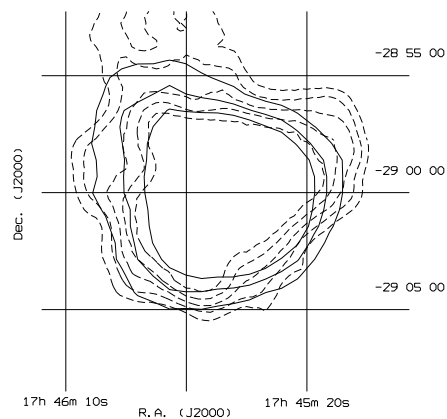,width=6.0cm} {\hfil}}
\caption{Contour levels of the  2--5 keV emission overlayed on the 90 cm radio 
contours (dashed line, adapted from  Anantharamaiah et al. 1991).}
\label{radio}
\end{figure}

Our observation of the SgrA complex confirms the presence of a hot 
plasma with multiple temperatures and/or in non-equilibrium ionization,
as already found with the ASCA   instruments (Koyama et al. 1996).
The increase in the 6.4 keV line equivalent width that we find 
in the North-East sector, is in agreement with the more detailed maps
of this line obtained with ASCA that show a correlation with the
molecular clouds.
The limited spectral resolution of the MECS, compared with that
of the ASCA solid state detectors, does not allow to study in more detail
the energy profile and spatial distribution of the individual lines.
On the other hand, 
the regular point spread function provided by the
BeppoSAX mirrors has allowed us to produce unbiased maps in wide energy bands
that demonstrate a clear difference in the spatial distribution of 
the softer and harder X--ray emission. 
Since also our spectral data could be well described by the sum of two 
thermal models with  kT$\sim$0.6 and 8 keV, it is tempting
to give an interpretation in terms of two plasma components   at 
different temperatures and with different spatial distributions
(although in reality the situation is 
certainly much more complex, with,
e.g., a distribution of temperatures). 
In this interpretation, it is remarkable that the lower temperature
plasma is well correlated with the   
SgrA~East triangular radio halo (Fig.\ref{radio}).

The presence of unresolved point sources could affect the
apparent distribution of the diffuse emission. We note however that previous
observations, e.g. with Einstein 
and ROSAT (Watson et al. 1981, Predehl \& Tr\"umper 1994) do not
show the presence of strong sources distributed 
in such a way to reproduce 
the triangular shape visible in  our low energy map. 
In particular,  the straight contours of the soft emission
corresponding to the SW side of the triangular radio halo
seem hardly explainable by a distribution of sources. Although 
we cannot exclude that  the apparent shape of the
softer X--rays is   due to an absorption effect, we favour 
the interpretation of
the lower temperature plasma   as a component physically 
related to the halo of SgrA~East.
In the following we will adopt this working hypothesis and  
assume a distance of 8.5 kpc.

We derive for the soft 
component an 
emission measure $EM$=(1.2$\pm{0.4})\times10^{14}$~cm$^{-5}$, which 
assuming emission from a spherical region with radius 10 pc   
and an electron filling factor $f$,  corresponds to  
n$_{e}\sim(3$~$\times$$f^{-1/2})$~cm$^{-3}$ and to a total
mass $M_{g}\sim250~M_{\odot}$$f^{1/2}$.
 
The average thermal pressure  
in the SgrA~East halo,  $P\sim3\times10^{-9}$~ergs~cm$^{-3}$, 
is consistent with the 
pressure $P_{Sedov}=0.106\times$$(E_{SN}/R^{3})$~ergs~cm$^{-3}$
derived  for a SNR in a Sedov phase, 
where $E_{SN}$ is the explosion energy of the SNR
and R is the shell radius in cm. Indeed, if we assume 
$E_{SN}=10^{51}$~ergs~s$^{-1}$ and 
R=10~pc, we find $P_{Sedov}\sim4\times10^{-9}$~ergs~cm$^{-3}$.

The   X--ray luminosity (2--10 keV) of the soft component
is $\sim4.5\times10^{35}$~ergs~s$^{-1}$. If we assume that this
thermal emission is mostly produced by the SgrA~East halo, 
its X--ray luminosity, pressure, density, 
temperature of the emitting gas (0.6 keV) 
and size ($\sim$20 pc), match well
with a supernova remnant origin in which thermal line emission
is produced when the expanding supernova
ejecta heats the ISM to X--ray temperatures.  

Another test of the SNR hypothesis can be done by comparing the
X--ray and radio surface brightnesses: indeed a strong 
correlation $\Sigma_{R}\propto\Sigma_{X}^{0.69\pm0.08}$
between the radio $\Sigma_{R}$ (at 1 GHz) and the X--ray surface 
brightness $\Sigma_{X}$ (0.15--4.5 keV)
was found by Berkhuijsen (1986). 
From the  radio spectrum reported by Pedlar et al. (1989)
we derived  
$\Sigma_{R}\sim5\times10^{-19}$~W~m$^{-2}$~Hz$^{-1}$~sr$^{-1}$. 
The $\Sigma_{X}$ measured from our X--ray data and converted  
to the 0.15--4.5 energy band  
is  $\Sigma_{X}\sim10^{34}$~ergs~s$^{-1}$~pc$^{-2}$.
These values fall well inside the region 
defined by other typical  SNRs.

In conclusion, all the physical quantities derived from 
the  analysis of the MECS are consistent with a
SNR origin for the SgrA~East halo.

\section{Conclusions}

The analysis of the diffuse
emission from the GC region reveals the presence 
of at least 
two distinct components : a 
soft one (T$_{mekal}\sim$0.6 keV),  
spatially correlated with the SgrA~East radio halo and a hard 
component (T$_{mekal}\sim$8--9 keV). 
While the hard component  is   elongated along
the   Galactic Plane and   does not spatially correlate with physical  
structures observed at other wavelengths, the spatial 
distribution of the soft component is remarkably well correlated
with the triangular radio halo of SgrA~East.
The parameters derived for the softer component
match extremely well with the idea that this thermal X--ray 
emission is due to a SNR origin
for the  SgrA~East  halo.

\begin{acknowledgements}

We thank L.~Chiappetti, G.~Matt, S.~Molendi and A.~Treves 
for useful discussions and help with the data analysis.

\end{acknowledgements}

\end{document}